\documentclass[preprint,prc,a4paper,tightenlines,nofootinbib]{revtex4}
\usepackage{epsfig,epsf,graphics,psfrag}
\usepackage{times}
\usepackage{color}
\usepackage{amsfonts,amssymb,stmaryrd,latexsym,amsmath}
\usepackage{hhline}

\newcommand{\beq}{\begin{equation}}
\newcommand{\eeq}{\end{equation}}
\newcommand{\beqa}{\begin{eqnarray}}
\newcommand{\eeqa}{\end{eqnarray}}
\newcommand{\nn}{\nonumber \\ }
\newcommand{\fet}[1]{\mbox{\boldmath $#1$}}

\begin{document} 

\title{Finite volume effects in low-energy neutron-deuteron scattering}

\bigskip\bigskip

\author{A.~Rokash}
\affiliation{Institut f\"ur Theoretische Physik II, Ruhr-Universit\"at Bochum, D-44g780 Bochum, Germany}

\author{E.~Epelbaum}
\affiliation{Institut f\"ur Theoretische Physik II, Ruhr-Universit\"at Bochum, D-44780 Bochum, Germany}

\author{H.~Krebs}
\affiliation{Institut f\"ur Theoretische Physik II, Ruhr-Universit\"at Bochum, D-44780 Bochum, Germany}


\author{D.~Lee}
\affiliation{Department of Physics, North Carolina State University,
  Raleigh, NC 27695, USA}

\author{U.-G.~Mei{\ss}ner}
\affiliation{Helmholtz Institut f\"{u}r Strahlen- und Kernphysik and
  Bethe Center for
Theoretical Physics, Universit\"{a}t Bonn,  D-53115 Bonn, Germany, and\\
Institut~f\"{u}r Kernphysik, Institute for Advanced Simulation,
J\"{u}lich Center for Hadron Physics,
Forschungszentrum J\"{u}lich, D-52425 J\"{u}lich, Germany}


\begin{abstract}
\noindent
We present a lattice calculation of neutron-deuteron scattering at very low energies and 
investigate in detail the impact of the topological finite-volume
corrections. 
Our calculations are carried out in the framework of pionless effective 
field theory at leading order in the low-energy expansion. Using lattice sizes
and a lattice spacing comparable 
to those employed in nuclear lattice simulations, we find that the
topological volume corrections must be taken into account  in order 
to obtain correct results for the neutron-proton S-wave scattering
lengths.   
\end{abstract}


\maketitle

\section{Introduction}
The so-called L\"uscher formula \cite{Lue1,Lue2} is well known to be a
standard tool in lattice calculations as it provides  
a relation between two-body energy levels in a finite 
box with periodical boundary conditions  and  phase shifts in the continuum. 
When describing scattering of composite particles such as e.g.~the deuteron 
on a lattice,  one has to take into account finite volume corrections to the
binding energy of a composite particle  in order to properly define the
scattering energy. These corrections are frame-dependent and topological 
in origin. For non-relativistic systems, the universal dependence of the
finite volume corrections to the binding energy on the momentum has been 
worked out in Ref.~\cite{Bour:2011ef}, see also Ref.~\cite{Davoudi:2011md}
for a generalization from non-relativistic quantum mechanics to quantum field 
theory. Here and in what follows, we will refer to this kind of corrections as 
the {\em topological volume corrections}.  

The impact of the topological volume corrections on the extraction of the
scattering parameters on the lattice  was studied in
Refs.~\cite{Bour:2011ef,Bour:2012hn} for the case of the three-body system consisting of equal mass 
two-component fermions in the universal shallow-binding limit. In this regime, the only relevant momentum 
scale is given by the dimer binding momentum $\kappa$ so that the quantity $\kappa a$, with $a$ 
referring to the fermion-dimer scattering length, represents a universal constant,  $\kappa a \sim 1.18$ 
\cite{Skor:1957,Stecher:2008,Levinsen:2011,Tan:2008} (note that the earlier
work of Ref.~\cite{Rupak:2006jj} found a slightly smaller value). 
The authors of Ref.~\cite{Bour:2011ef} found the inclusion of the topological 
volume corrections to be essential for  obtaining the correct continuum
limit. The effect was especially pronounced for the effective range which 
is known to be rather small in magnitude. 

In the present work,  we investigate the role of the topological volume
corrections in a realistic system, namely neutron-deuteron scattering at very low energy.
The calculations are carried out in the framework of pionless effective 
field theory (EFT) at leading order in the low-energy expansion. Specifically, we calculate the 
neutron-deuteron S-wave scattering lengths in both the spin-doublet and spin-quartet channels 
on the lattice with and without taking into account the topological volume
corrections. In order to draw conclusions, 
we compare our lattice results with  the continuum ones emerging from solving the 
Skorniakov-Ter-Martirosian (STM) equation using the same input parameters~\cite{Skor:1957}. 

Our paper is organized as follows. In section~\ref{sec2}, we briefly
discuss  the topological volume corrections 
following  Ref.~\cite{Bour:2011ef}. The description of nucleon-deuteron 
scattering at very low energy based on the STM integral equation is briefly 
reviewed in section~\ref{sec3}. A detailed description of 
the lattice calculations is given in section~\ref{sec4} while the main 
findings of our study are summarized in section~\ref{sec5}.

\section{Topological volume corrections}
\label{sec2}

As already pointed out in the introduction, the standard tool to extract 
phase shifts on the lattice is the finite volume formula derived by  
L\"uscher \cite{Lue1,Lue2} which relates the energy 
levels of a two-body system in a cubic periodic box of length $L$ to the scattering phase shifts.  
For the $S$-wave case we are interested in, this relation has the form
\begin{equation}
p \cot \delta = \frac{1}{\pi L} S(\eta), 
\quad\quad
\eta \equiv \left(\frac{pL}{2\pi}\right)^2,
\label{pcotdelta}
\end{equation}
where the three-dimensional zeta function $S(\eta)$ is given by
\begin{equation}
S(\eta)=\lim_{\Lambda\rightarrow\infty}\left[  \sum_{\vec{k}\in \mathbb{Z}^3}\frac
{\theta(\Lambda^{2}-\vec{k}^{2})}{\vec{k}^{2}-\eta}-4\pi\Lambda\right].
\end{equation}
The only input parameters entering L\"uscher's formula in
Eq.~(\ref{pcotdelta}) are the relative momentum  $\vec{p}$
of the scattered particles and the box size $L$. For point-like particles, 
the relative momentum $p \equiv | \vec p \, |$ can be directly
inferred from the energy spectrum measured on the lattice. 
For composite particles, however, the total energy of the system receives contributions from 
the binding energy of the scattered particles in addition to their kinetic
energies. In a finite volume, one needs to account for topological 
corrections emerging from the bound state wave function touching all boundaries of the box. These corrections 
have to be subtracted from the total energy of the system in order to
correctly determine the value of the relative 
momentum $p$ to be inserted in Eq.~(\ref{pcotdelta}).  We now briefly outline 
the derivation of these corrections following closely  Refs.~\cite{Bour:2011ef,Bour:2012hn}. 

We start with the scattering wave function corresponding to the momentum $p$ that enters the 
derivation of Eq.~(\ref{pcotdelta}). Outside the interaction region,
it  can be written in the form 
\begin{eqnarray}
\label{perturb2}
\langle \vec r \,|\Psi_p \rangle \propto\sum\limits_{\vec{k}\in \mathbb{Z}^3}
\frac{\mathrm{e}^{2\pi i\vec{k}\cdot\vec{r}/L}}{\vec{k}^2- \eta}\ .
\end{eqnarray}
The corresponding total energy of the system in the case of two composite particles $A$ and $B$ is given by 
\begin{eqnarray}
\label{perturb5}
E_{AB}(\eta ,L)=\frac{\langle \Psi_p | H | \Psi_p \rangle }{\langle
  \Psi_p | \Psi_p \rangle } =
\Bigg[\sum\limits_{\vec{k}\in \mathbb{Z}^3}\frac{1}{\big(\vec{k}^2-\eta \big)^2}\Bigg]^{-1}\!\!\!\!\sum\limits_{\vec{k}\in \mathbb{Z}^3}\frac{\frac{\vec{p}^2}{2 m}+E^A_{\vec{k}}(L)+E^B_{\vec{k}}(L)}{\big(\vec{k}^2- \eta \big)^2}\,,
\end{eqnarray}
where $E^A_{\vec{k}}(L)$ and $E^A_{\vec{k}}(L)$ are the binding energies of
the particles $A$ and $B$ moving with momentum $2 \pi \vec k/L$, and $m$ is 
their reduced mass.  In order to evaluate the sum in the above equation, one needs to relate 
the finite-volume binding energy corrections for $\vec k \neq \vec 0$ to that
in the rest--frame of the particle corresponding to 
$\vec k = \vec 0$. 

Consider first the wave function $\phi_\infty (\vec r \, )$ of the two-body
bound state $A$ or $B$ at rest in the infinite volume. 
The relative coordinate $\vec r $ is defined in terms of the coordinates 
$\vec r_1$ and $\vec r_2$ of the constituents 
$1$ and $2$ as  $\vec r = \vec r_1 - \vec r_2$.   
As shown by L\"uscher in Ref.~\cite{Lue1}, see also
Refs.~\cite{Konig:2011nz,Konig:2011ti}, 
the finite-volume correction to the binding energy of 
two-body states with $\vec k = \vec 0$, bound by a short range potential 
$V(\vec r \, )$\footnote{We assume that the range of the potential is much
  smaller than the lattice size $L$.}, is given by 
\begin{eqnarray}
\label{luescherbound}
\Delta E_{\vec 0} (L) \equiv E(L)-E(\infty) = \sum_{|\vec{n}|=1}\int{\mathrm{d^3r} \, \phi_{\infty}^{*}(\vec{r})V(\vec r \, )\phi_{\infty}(\vec{r}+\vec{n}L)}
+\mathcal{O}\left(\mathrm{e}^{-\sqrt{2}\kappa L}\right) \,.
\end{eqnarray}
Evaluating the right-hand-side of this equation for S-wave bound states 
yields L\"uscher's result
\begin{equation}
\Delta E_{\vec 0} (L) =  - 3 | \gamma |^2 \frac{\mathrm{e}^{-\kappa L}}{\mu L} 
+\mathcal{O}\left(\mathrm{e}^{-\sqrt{2}\kappa L}\right) \,,
\end{equation}
where $\mu$ and $\gamma$ denote the reduced mass and the asymptotic wave function normalization. 
For a comprehensive derivation of this result and its generalization
to higher partial waves and to the case of 
two spatial dimensions see Ref.~\cite{Konig:2011ti}. 

For a bound state moving with the momentum $\vec{P}$ in an infinite volume, the part 
of the wave function which depends on the center-of-mass coordinate $\vec{R}$ is simply a plane wave. 
Introducing a finite-volume box of the size $L$, imposing periodic boundary conditions and 
parametrizing the total momentum $\vec P$ in terms of an integer  
vector $\vec k$ via $\vec{P}=(2\pi/ L)\vec{k}$, one obtains the relation 
\begin{eqnarray}\label{period3}
\phi_L(\vec{r} \, )=\mathrm{e}^{i2\pi\alpha\vec{k}\cdot\vec{n}}\phi_L(\vec{r}+\vec{n}L), \qquad \textnormal{where} \ \ \vec{k},\vec{n} \in \mathbb{Z}^3 \,
\end{eqnarray}
where $\alpha=m_1/(m_1+m_2)$ and $m_1$, $m_2$ denote the masses of the constituents $1$, $2$, respectively.   
The physical meaning of this
twisted boundary condition for $\phi_L(\vec{r})$ becomes obvious by observing that the phase generated 
by the part of the wave function associated with the 
center-of-mass motion has to be absorbed by the wave function $\phi_L(\vec{r}) $ describing the
relative motion of the two constituents. Using the twisted boundary condition in Eq.~(\ref{period3})
when evaluating the integral in the right-hand-side of Eq.~(\ref{luescherbound}) allows one to derive 
a universal relation between the finite-volume binding energy correction for an S-wave two-body 
state moving with the momentum $\vec P = (2\pi/ L)\vec{k}$,  $\Delta E_{\vec k}$, to the one $\Delta E_{\vec 0}$
for a state at rest:   
\begin{eqnarray}\label{luescher3}
\frac{\Delta E_{\vec{k}}(L)}{\Delta E_{\vec 0}(L)}=\frac{1}{3}\sum\limits_{l=1}^{3}\cos(2\pi\alpha k_l)+\mathcal{O}\left(\mathrm{e}^{-\kappa L}\right)\,,
\end{eqnarray}
see Ref.~\cite{Konig:2011ti} for more details. 
Finally, combining Eqs.~(\ref{luescher3}) and (\ref{perturb5}), one obtains
\begin{eqnarray}
\label{perturb7}
E_{AB}(\eta ,L)-E_{AB}(\eta,\infty)=\Delta E^A_{\vec 0}(L)T( \eta ,\alpha_A)+\Delta E^B_{\vec 0}(L)T( \eta ,\alpha_B)
\end{eqnarray}
where the quantity $T$ is defined as   
\begin{eqnarray}
\label{topological volume factor2}
T(\eta,\alpha)=\Bigg(\sum\limits_{\vec{k}\in \mathbb{Z}^3}\frac{1}{\big(\vec{k}^2-\eta \big)^2}\Bigg)^{-1}\sum\limits_{\vec{k}\in \mathbb{Z}^3}\frac{\frac{1}{3}\sum\limits_{l=1}^{3}\cos(2\pi\alpha k_l)}{\big(\vec{k}^2- \eta \big)^2}\,. 
\end{eqnarray}
To summarize, the proper determination of the relative momentum $p$ or, equivalently, 
the corresponding dimensionless quantity $\eta$ from the energy spectrum measured on the lattice requires 
solving Eq.~(\ref{perturb5}). The goal of our work is to demonstrate the importance of the topological 
volume correction in the case of composite particles given in Eq.~(\ref{perturb7}) for precision 
determination of low-energy neutron-deuteron scattering parameters in lattice EFT simulations.

\section{Low-energy neutron-deuteron scattering in the continuum}
\label{sec3}

Low-energy neutron-deuteron scattering is extensively studied in the framework 
of both chiral and pionless EFT, see \cite{Epelbaum:2008ga,Hammer:2010kp} for recent review articles
and references therein.  For the purpose of the present analysis aiming at a precision benchmark 
calculation of the S-wave neutron-deuteron scattering lengths, we restrict ourselves to the 
simplest possible formulation of pionless EFT at lowest order \cite{Bedaque:1999ve}. We now briefly 
outline the framework and provide the relevant results.   

The lowest-order effective Lagrangian in pionless EFT can be written as 
\begin{eqnarray}
\label{lagr}
\mathcal{L} &=& N^\dagger \Bigg( i \partial_0 + \frac{\vec \nabla^2}{2 m} \Bigg) N 
- \frac{1}{2} C_0 (N^\dagger N)  (N^\dagger N)   
- \frac{1}{2} C_I (N^\dagger \fet \tau N) \cdot  (N^\dagger \fet \tau N)  \nonumber \\
&-& 
\frac{1}{6} D (N^\dagger N)  (N^\dagger N)   (N^\dagger N) + \ldots \,, 
\end{eqnarray}
where $N$ denote the fields associated with non-relativistic nucleons, the 
ellipsis represent higher-order terms 
involving derivatives and  $\fet \tau$ refer to 
the Pauli matrices acting in the isospin space. Further, $m$ is the nucleon 
mass\footnote{Here and in what follows, we work in the exact isospin
  limit.} while $C_{0}$,  $C_{I}$ and 
$D_{0}$ are the relevant low-energy constants (LECs).  Notice that because of
the antisymmetric nature of the  fermionic states, there 
are only two independent derivative-less two-nucleon (2N) interaction and only
one derivative-less three-nucleon  (3N) contact force, see \cite{Bedaque:1999ve} for more details.  

Pionless EFT allows one to compute few-nucleon observables at momenta $Q$ well below the pion mass
$M_\pi$  representing the breakdown scale of this approach.  
In the 2N sector, the effective Lagrangian in Eq.~(\ref{lagr}) gives rise to
the S-wave scattering amplitude 
\beq
T_2(p) = \frac{4 \pi}{m} \; \frac{1}{-1/a_2 - i p} + \ldots
\eeq
where $\vec p$ is the nucleon momentum in the center-of-mass system and $a_2$ is an S-wave 
scattering length. Notice that, given the fact that the scattering lengths in the spin-singlet (i.e.~$^1S_0$) and  
spin-triplet (i.e.~$^3S_1$) neutron-proton channels are rather large in
magnitude, $a_2^s = -23.7$ fm and 
 $a_2^t = 5.4$ fm, respectively, it is necessary to non-perturbatively resum 
the $C_{0,I}$-interactions  in order 
for the results to be applicable in the range of momenta $| a_2|^{-1} \ll Q \ll M_\pi$.    

The neutron-deuteron ($nd$) scattering amplitude can be most conveniently 
calculated  by rewriting the theory  in terms of the so-called ``dimeron'' 
auxiliary fields \cite{Kaplan:1996nv} which couple to the two-nucleon states in 
the $^1S_0$ and  $^3S_1$ channels. Using the dimeron field allows one to get rid of the 
$C_{0,I}$ interactions in the Lagrangian and to reduce the system of Faddeev equations describing 
three-nucleon scattering to much simpler integral equations for nucleon-dimeron scattering, 
see \cite{Bedaque:1999ve} for more details. In particular, for the
spin-quartet $nd$ channel, one obtains the 
STM equation  for the scattering amplitude $a (p,k)$
\cite{Skor:1957,Bedaque:1997qi}  
\begin{eqnarray}
\label{skor2}
\frac{3}{4}\bigg(\frac{1}{a_2^t}+\sqrt{3p^2/4-mE}\bigg)^{-1}a(p,k)
= -K(p,k)-\frac{2}{\pi}\int\limits_0^{\infty}\frac{q^2\mathrm{dq}}{q^2-k^2-i\epsilon}
K(p,q)a(q,k)\,,
\end{eqnarray}
where $k$ and $p$ denote the incoming and outgoing momenta in the
center-of-mass system, $E = (3 k^2/4 - 1/(a_2^t)^2)/m$  is the total 
energy and  the kernel $K$ is given by    
\begin{equation}
 K(p,q)=\frac{1}{2pq}\ln\left(\frac{q^2+pq-p^2-mE}{q^2-pq-p^2-mE}\right)\, .
\end{equation}
Notice that the derivative-less 3N contact interaction does not contribute 
to the spin-quartet channel due to the Pauli principle.   
In Eq.~(\ref{skor2}), the relevant linear combination of the LECs $C_0 - 3 C_I$  is 
expressed in terms of the NN spin-triplet scattering length $a_2^t$.  
Expanding the amplitude $a(k,k)$ around $k =0$ then yields a prediction for the 
spin-quartet $nd$ scattering length $a_3^q$.  Using the deuteron
binding energy $B_d = 2.22$~MeV to fix the 2N contact interaction, 
which at lowest order of EFT corresponds to $a_2^t= 1/\sqrt{m B_d} 
= 4.32$ fm, one obtains a prediction for the neutron-deuteron quartet 
scattering length  \cite{Skor:1957,Bedaque:1997qi}    
\beq
\label{benchq}
a_3^q \simeq 5.1 \mbox{ fm}\,,
\eeq
which is in a satisfactory agreement with the experimental value of $(a_3^q )_{\rm
  exp} = 6.35 \pm 0.02$ fm \cite{Dilg:1971}.  
It is also well known that the agreement between the theory and experiment 
is strongly improved by taking  into account corrections due to the effective
range in the two-body system at higher orders of the EFT expansion
\cite{Bedaque:1997qi}. 

For the spin-doublet $nd$ channel, the two-nucleon subsystem can be both in
the $^1S_0$ and $^3S_1$ channels.  Consequently, one obtains a system of
coupled STM-like  equations \cite{Skor:1957}
which in the notation of Ref.~\cite{Bedaque:1999ve} has the form   
\begin{eqnarray}
 \frac{3}{2}\Bigg(\frac{1}{a_2^t}+\sqrt{\frac{3p^2}{4}-mE}\Bigg)^{-1}a(p,k)&=&K(p,k)+\frac{2H(\Lambda)}{\Lambda^2}
+\frac{2}{\pi}\int\limits_0^{\Lambda}\frac{q^2\mathrm{dq}}{q^2-k^2-i\epsilon}\nn
&\times &   \Bigg[K(p,q)(a(q,k) + 3b(q,k))+
\frac{2H(\Lambda)}{\Lambda^2}(a(q,k)+b(q,k))\Bigg],\nonumber\\
2\frac{\sqrt{3p^2/4-mE}-1/a_2^s}{p^2-k^2}\; b(p,k)&=&3K(p,k)+\frac{2H(\Lambda)}{\Lambda^2}
+\frac{2}{\pi}\int\limits_0^{\Lambda}\frac{q^2\mathrm{dq}}{q^2-k^2-i\epsilon}\nn
&\times & \Bigg[K(p,q)(3a(q,k) +b(q,k))+
\frac{2H(\Lambda)}{\Lambda^2}(a(q,k)+b(q,k))\Bigg]\,,\nonumber
\end{eqnarray}
where $a (p,k)$ ($b(p,k)$) is the spin-doublet nucleon-deuteron
scattering amplitude (the amplitude describing a transition to a spin-$0$,
isospin-$1$ dibaryon). Further, $H (\Lambda ) \propto D$ parameterizes the 
3N force. Contrary to the spin-quartet channel, the
coupled integral equations do not possess a unique solution in the
$\Lambda \to \infty$ limit in the absence of the 3N force which is due
to the fact that the corresponding homogeneous equations have a 
non-trivial solution, see \cite{Danilov:1963}
for more details.  When solving the equations with a finite cutoff
$\Lambda$, this results in a strong $\Lambda$-dependence of the
scattering amplitudes. The  cutoff dependence is absorbed into
the ``running'' of the 3N force $H(\Lambda )$. Notice that the
functional dependence of $H$ of $\Lambda$ is known and can be found
e.g. in Ref.~\cite{Bedaque:1998kg}. One needs a single 3N datum to fix the value of
$H$, which allows one then to make predictions for all other low-energy observables in 
this channel. This leads, in particular, to the well-known correlation
between the triton binding energy and the $nd$ doublet S-wave scattering
length $a_3^d$, the so-called Phillips line \cite{Phillips:1968zze}.\footnote{Notice that this
correlation is broken by higher-order corrections so that the Phillips
line actually turns into a band.} Using the deuteron and triton
binding energies $B_d = 2.22$~MeV and $B_t = 8.48$~MeV  together with 
the $np$ $^1S_0$ scattering length  $a_2^s =-23.7$~fm, the resulting value for the  
$nd$ spin-doublet scattering length $a_3^d$ is \cite{Bedaque:1999ve}
\beq 
\label{benchd}
a_3^d\simeq 0.5 \mbox{ fm}\,.
\eeq
It is also rather close to the experimental value of $(a_3^d )_{\rm exp} 
= 0.65 \pm 0.04$~fm \cite{Dilg:1971}.  
The values of the $nd$ scattering lengths quoted in
Eqs.~(\ref{benchq}) and (\ref{benchd}) will serve as reference points for
our lattice calculations.

\section{Neutron-deuteron scattering on the lattice}
\label{sec4}

We now turn to the discretized version of pionless EFT and carry out a precision
determination of the $n$d scattering lengths $a_3^q$ and $a_3^d$ on the
lattice. We restrict ourselves to lowest order and use the
same input as described in the previous section. Note that a first
investigation of the 3N system in pionless EFT on the lattice was already 
given in Ref.~\cite{Borasoy:2005yc}.

\subsection{Lattice Hamiltonian}

We first specify the lattice notation and the discretized form of the
Hamiltonian corresponding to the Lagrangian in Eq.~(\ref{lagr})
following the lines of Ref.~\cite{Lee:2008fa}. We define the creation operator 
$a^{\dagger}_{0}(\vec{r}\, )=a^{\dagger}_{n}(\vec{r} \, )$ for the neutron, and
$a^{\dagger}_{1}(\vec{r} \, )=a^{\dagger}_{p}(\vec{r} \, )$ for the proton.  
To shorten the notation, it is convenient to introduce the nucleon density
operators $\rho^{a^{\dagger},a}(\vec{r} \, )$ and
$\rho^{a^{\dagger},a}_I(\vec{r} \, )$ according to 
\begin{equation}
 \rho^{a^{\dagger},a}(\vec{r} \, )=\sum_{i=0,1}a^{\dagger}_{i}(\vec{r}
 \, )a_{i}(\vec{r} \, )
 \,, \quad \quad 
 \rho^{a^{\dagger},a}_I(\vec{r}\,
 )=\sum_{i,i'=0,1}a^{\dagger}_{i}(\vec{r} \,
 )\left[\tau_I\right]_{ii'}a_{i'}(\vec{r} \, ) \,.
\end{equation}
The leading order effective Hamiltonian can then be written as
\begin{eqnarray}
 H=H_0+V_{2N}+V_{3N}, 
\end{eqnarray}
where $H_0$ is the free Hamiltonian
\begin{eqnarray}
 H_0=-\frac{1}{2m}\sum_{i}\int\mathrm{d}^3\vec{r}a_i^{\dagger}(\vec{r}
 \, )\Delta a_i(\vec{r} \, ), 
\end{eqnarray}
while the 2N and 3N contact interactions have the form 
\begin{eqnarray}
V_{2N} &=&
\frac{C_0}{2}\int\mathrm{d}^3\vec{r}:\left[\rho^{a^{\dagger},a}(\vec{r}
  \, )\right]^2:
+
\frac{C_{I^2}}{2}\sum_{I}\int\mathrm{d}^3\vec{r}:\left[\rho_I^{a^{\dagger},a}(\vec{r}
  \, )\right]^2:
\,, \nn
V_{3N}&=&
\frac{D}{6}\int\mathrm{d}^3\vec{r}:\left[\rho^{a^{\dagger},a}(\vec{r}
  \, )\right]^3:\,,
\end{eqnarray}
where $:\, :$ indicates that the operators are taken in normal ordering. 
Here and in what follows, we employ an $\mathcal{O}(a^2)$-improved
lattice approximation for the Laplacian yielding the discretized free Hamiltonian
in the form 
\begin{eqnarray}
 H_0&=&\frac{1}{2m}\sum_{i,\vec{n}}\sum_{\mu=e_x,e_y,e_z}\big[ \frac{5}{2}a^{\dagger}_{i}(\vec{n})a_{i}(\vec{n})-\frac{4}{3}
\left(a^{\dagger}_{i}(\vec{n})a_{i}(\vec{n}+\mu)+a^{\dagger}_{i}(\vec{n})a_{i}(\vec{n}-\mu)\right)\nonumber\\
&+&\frac{1}{12}\left(a^{\dagger}_{i}(\vec{n})a_{i}(\vec{n}+2\mu)+a^{\dagger}_{i}(\vec{n})a_{i}(\vec{n}-2\mu)\right)\big].
\end{eqnarray}

\subsection{Determination of the low-energy constants}
\label{secLEC}

\begin{figure}[t]
\centering
\includegraphics[width=0.8\textwidth]{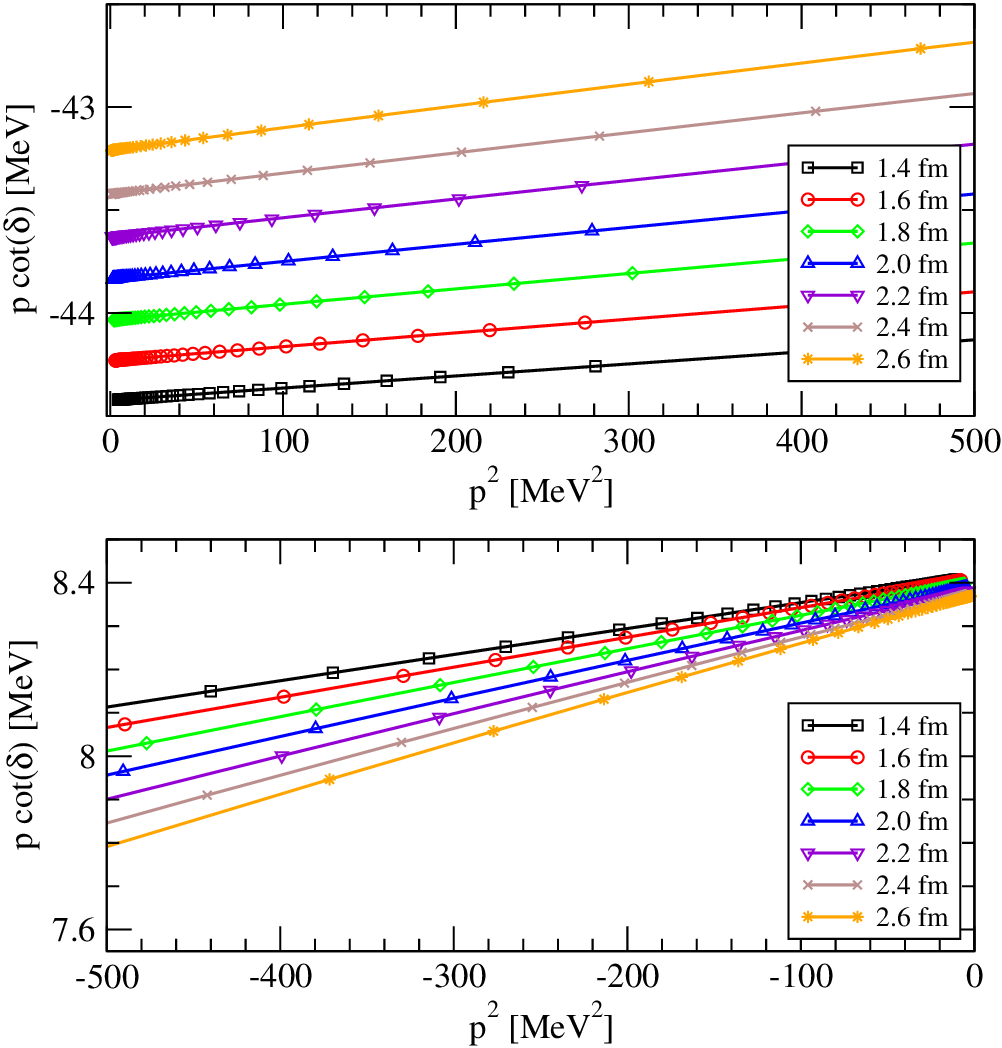}
\caption{Effective range function in the $^3S_1$ 
(top panel) and $^1S_0$ (bottom panel)  neutron-proton  channels.   
}
\label{111}	
\end{figure}

We need three independent conditions to fix the low-energy constants
(LECs) $C_0$, $C_{I^2}$ and
$D$. In the two-body sector, we tune the constants $C_0$ and $C_{I^2}$
such that the ground state in the $^3S_1$ channel has the energy of
the deuteron, $E_d=-2.22$ MeV, and the experimental value of the 
scattering length in the $^1S_0$ channel, $a_2^s=-23.7$ fm, is reproduced. 
The energy spectrum of the two-body system is determined by
diagonalizing the Hamiltonian on the lattice. Here and in what
follows, we employ the spatial lattice spacings in the range $a_{\rm latt} = 1.4
\ldots 2.6$~fm which are also typical for chiral EFT nuclear lattice simulations of 
light nuclei
\cite{Epelbaum:2009zsa,Epelbaum:2009pd,Epelbaum:2010xt,Epelbaum:2011md},
dilute neutron matter \cite{Epelbaum:2008vj}, and to the structure of the 
Hoyle state~\cite{Epelbaum:2012qn}. To calculate the
$^1S_0$ scattering length from the energy spectrum, we first use the
L\"uscher formula, see Eq.~(\ref{pcotdelta}), to obtain the effective
range function $p \cot (\delta )$ as depicted in Fig.~\ref{111}, where 
we also show the results in the spin-triplet channel.  
The corresponding values of the scattering length and effective
range can then be easily determined by a polynomial
fit. Our results for
the LECs $C_0$ and $C_{I^2}$ for various lattice spacings are summarized in 
Tab.~\ref{coefficients} in lattice units. 
\begin{table}[b]
\begin{center}
\begin{tabular*}{0.7\textwidth}{@{\extracolsep{\fill}}|c||c c c|}
\hline 
&&& \\ [-10pt]
   $a_{\rm latt}$ [fm]       & $C_0$ $\big[ a_{\rm latt}^{2} \big]$ & $C_{I^2}$
   $\big[ a_{\rm latt}^{2} \big]$ 
   & $D$ $\big[ a_{\rm latt}^{5} \big]$   \\ 
&&& \\ [-10pt]
 \hline \hline
 1.4&  $-0.6969$   & $0.0257$   &$0.9070 $\\
 1.6 & $-0.6115$ & $0.0260 $ &$0.7505$	\\
 1.8 & $-0.5450$   &$0.0264$  &$0.6344$ 		\\
 2.0&  $-0.4920$ & $0.0267$  &$0.5456$	\\
 2.2 &  $-0.4488$ &$0.0270$  &$0.4748 $\\
2.4 &  $-0.4128$ & $0.0273$ &$0.4168$ \\
2.6 &  $-0.3824$ &$0.0276 $ &$0.3683 $\\ \hline	
\end{tabular*}
\end{center}
\caption{Low-energy constants $C_0$, $C_{I^2}$ and
$D$  for  various values of the lattice spacing $a_{\rm latt}$.}
\label{coefficients}	
\end{table}
The much smaller absolute values for the LEC $C_{I^2}$ reflects
the approximate SU(4) Wigner symmetry of the 2N interactions 
\cite{Mehen:1999qs}.  

\begin{figure}[t!]
\vskip 2.5 true cm 
\centering
\includegraphics[width=1.\textwidth]{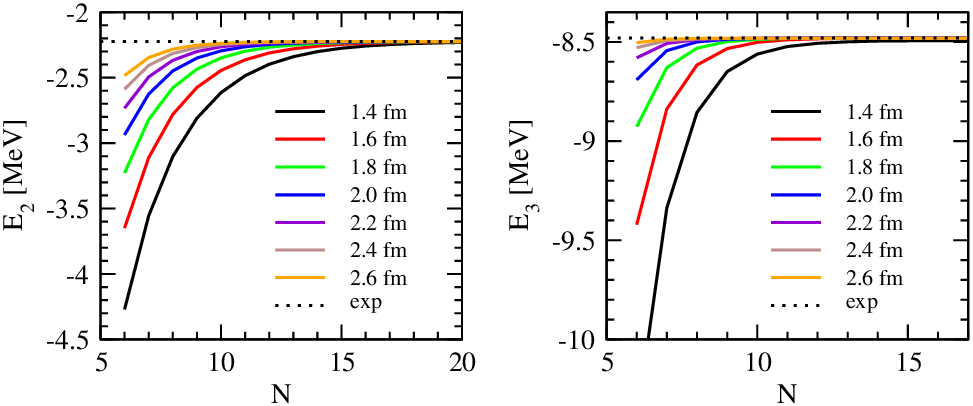}
\caption{Lattice size dependence of the dimer (left panel) and trimer
  (right panel) energies. 
}
\label{333}
\end{figure}

For the three-nucleon system, we use the
Lanczos method to determine the lowest eigenvalues of the
Hamiltonian. 
The LEC $D$ is determined by the requirement to reproduce the triton
energy $E_t=-8.48$~MeV. The resulting values of this LEC are listed in
Tab.~\ref{coefficients}.   
For both the deuteron (also called dimer) and the triton (also called 
trimer) binding energies, we do observe
sizable volume effects when using lattice sizes of the order of $L
= N a_{\rm latt}\sim 20$ fm and smaller (and varying the lattice spacing in the range
specified above).  This is visualized in Fig.~\ref{333}.  

Finally, we give in Tab.~\ref{otherranges} the infinite-volume values of the various two-body
scattering parameters not used in the fit, namely the spin-triplet
scattering length $a_2^t$ and  the effective range parameters $r_2^s$
and $r_2^t$ in the $np$ $^1S_0$ and $^3S_1$ channels, respectively. 
The non-vanishing values of the effective range parameters can be
traced back to appearance of the ultraviolet cutoff set in our
calculations by a finite lattice spacing. 
Extrapolating to the continuum we observe $r_2^{s,t} \simeq 0$ fm  and 
$a_2^t\approx 1/\sqrt{m B_d} = 4.32$ fm in agreement with an infinite-cutoff limit of
leading-order pionless EFT. 
\begin{table}[t]
\begin{center}
\begin{tabular*}{0.7\textwidth}{@{\extracolsep{\fill}}|c||c c c|}
\hline 
&&& \\ [-10pt]
   $a_{\rm latt}$ [fm]      & $a_2^t$ [fm] & $r_2^t$ [fm]  &$r_2^s$ [fm]   \\  
&&& \\ [-10pt]
 \hline \hline
 1.4&  $4.44$   &$0.23$&$0.23$ \\
 1.6 & $4.46$ &  $0.27$&$0.27$\\
 1.8 & $4.48$   & $0.30$&$0.30$ \\
 2.0&  $4.50$ &   $0.33$&$0.33$\\
 2.2 &  $4.52$ &  $0.37$&$0.37$ \\
2.4 &  $4.54$ &  $0.40$&$0.40$ \\
2.6 &  $4.57$ &  $0.43$&$0.44$ \\ \hline	
\end{tabular*}
\end{center}
\caption{Various two-body scattering parameters not used in the fit of
 the  LECs at various values of the lattice spacing $a_{\rm
   latt}$. All values correspond to the infinite volume. 
}
\label{otherranges}	
\end{table}

\subsection{Neutron-deuteron effective range function}

Consider now neutron-deuteron scattering.   Eq.~(\ref{perturb7}) for
the volume corrections to the energy of the system $E_{nd}$ reduces to 
\begin{eqnarray}\label{topol} 
E_{nd}(\eta , L) =E_{d}(\infty) + \frac{p^2}{2m_{\rm eff}}
+T(\eta,1/2)\left(E_{d}(L)-E_{d}(\infty)\right),
\end{eqnarray}
where we used the infinite-volume relation 
\begin{eqnarray} 
 E_{nd}(\eta,\infty)=\frac{p^2}{2m_{\rm eff}}+E_{d}(\infty)\,.
\end{eqnarray}
Here, $E_d(\infty)$ is the physical value of the
(negative) deuteron binding energy used as an input parameter
as described in Sec.~\ref{secLEC}.  
Further, the lattice value of the reduced mass of the nucleon-deuteron system $m_{\rm
  eff}$ is determined by measuring the deuteron dispersion relation for
a given value of the lattice spacing $a_{\rm latt}$.  

We remind the reader that the $nd$ relative momentum $p$ to be
determined is related to the quantity $\eta$ via $\eta =(p L/(2
\pi))^2$.  To determine the value of
$\eta$ which enters  the L\"uscher formula to
calculate the  $nd$ phase shifts, we solve Eq.~(\ref{topol}) iteratively using the method 
described in Ref.~\cite{Bour:2012hn}.  Since we are particularly
interested here in the effect of the topological volume corrections,
we also perform calculations with these corrections being switched
off, i.e.~ we set $T(\eta , 1/2) =0$. 
Eq.~(\ref{topol}) then reduces to
\begin{eqnarray}\label{topol0} 
E_{nd}(\eta, L)-E_{d}(\infty)=\frac{p^2}{2m_{\rm eff}}
\end{eqnarray}
and allows one to directly read off  the  momentum $p$.  

\begin{figure}[t]
\centering
\includegraphics[width=0.85\textwidth]{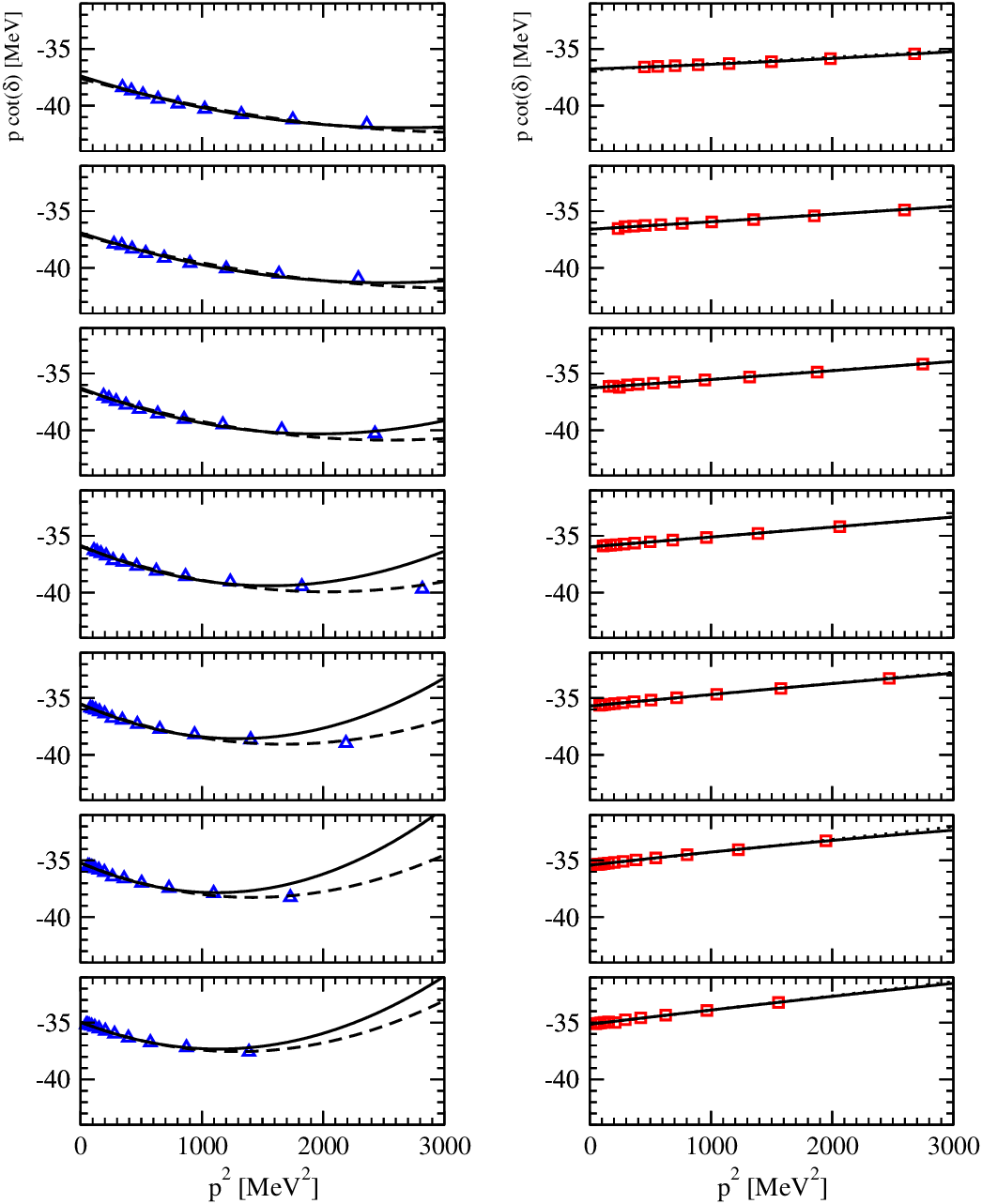}
\caption{Effective range function in  spin-quartet S-wave
  neutron-deuteron scattering for various values of the  lattice 
 spacings $a_{\rm latt}$. 
Left panel: without topological corrections; right panel: with
topological corrections. Solid (dashed) lines
refer to fits to all lattice points (all points except the one for
largest $p^2$).}
\label{444q}
\end{figure}

\begin{figure}[t]
\includegraphics[width=0.85\textwidth]{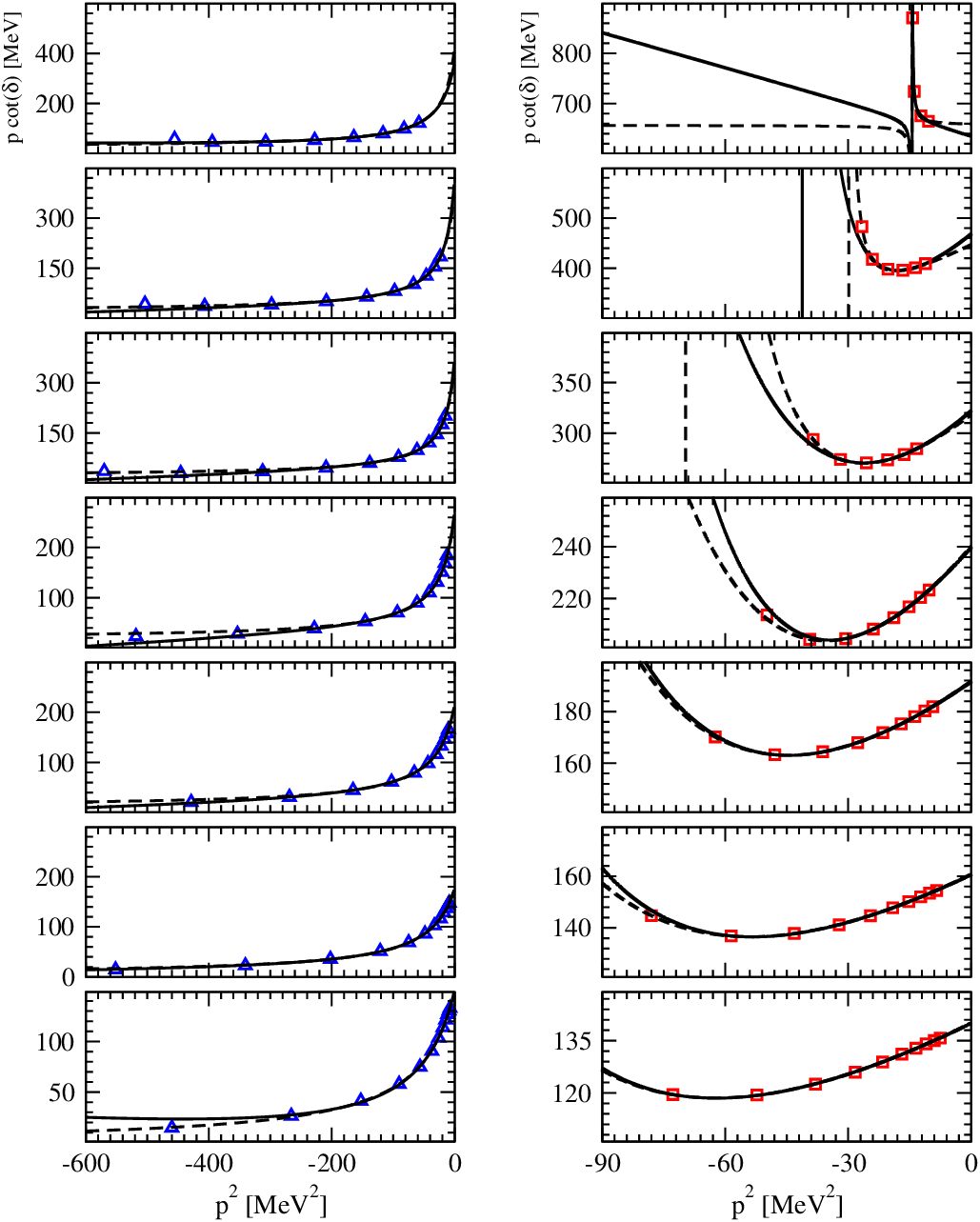}
\caption{Effective range function in  spin-doublet S-wave
  neutron-deuteron scattering for various values of the  lattice 
  spacings $a_{\rm latt}$. 
Left panel: without topological corrections; right panel: with
topological corrections. For further
notation, see Fig.~\ref{444q}.}
\label{444d}
\end{figure}

With the relative momentum $p$ being determined as described above, it
is straightforward to extract the phase shifts using
Eq.~(\ref{pcotdelta}).
We evaluate the spectrum of the three-body system for a different number of
lattice size $N=6\ldots 17$ to obtain a number of different values
of $p$ for each lattice spacing. 
The resulting effective range functions for the various 
values of $a_{\rm latt}$ are depicted in Fig.~\ref{444q}
for the spin-quartet channel and in Fig.~\ref{444d}
for the spin-doublet channel. 
While the function $p \cot (\delta )$ has a rather smooth behavior at
small $p$ in the quartet channel and can be well described in terms 
of the effective range expansion   
\beq
p \cot ( \delta ) = - \frac{1}{a} + \frac{1}{2} r_{\rm eff} p^2 + v_2 p^4
+ \ldots\,, 
\eeq
where $v_2$ is the first shape parameter, it shows a rather strong
curvature in the near-threshold region in the spin-doublet channel. 
This is in line with a well-known fact that the function $p \cot
(\delta )$ has a pole near the elastic threshold \cite{oers1967,Phillips:1977wh}. 
In order to be able to extract the scattering length, we employ a
parametrization due to van Oers and Seagrave \cite{oers1967} which accounts for
the appearance of this pole
\begin{eqnarray}\label{EREform}
p\cot(\delta)=-\frac{1}{a}\left(\frac{1+b_2p^2+b_4p^4+\ldots}{1+dp^2}\right),
\end{eqnarray}
where the coefficients $b_i$ and $d$ can, of course, be expressed in
terms of the effective range and shape parameters. 
Given that the results for large lattice sizes (and thus for small
values of $p$) are both more reliable and more appropriate for the
effective range expansion, we apply a weighting factor $p_i^{-1}$ when
carrying out the $\chi^2$ fits to extract the coefficients $a$, $b_i$ and
$d$ ($a$, $r_{\rm eff}$ and $v_i$) in the spin-doublet
(spin-quartet) channel. Our results for the scattering lengths  
are collected in Tab.~\ref{ScatteringLengthQuartetDoublet}.  
\begin{table}[t]
\begin{center}
\begin{tabular*}{0.9\textwidth}{@{\extracolsep{\fill}}|c||c c||c c|}
\hline
& \multicolumn{2}{c||}{spin-quartet}  & \multicolumn{2}{c|}{spin-doublet} \\
\hline \hline
   $a_{\rm latt}$ [fm]       & $(a_3^{q})^{nc}\; $ [fm]  & $(a_3^{q})^{c}\; $ [fm] & $(a_3^{d})^{nc}\; $ [fm]  & $(a_3^{d})^{c}\; $ [fm]   \\  \hline
 1.4&  $5.28\pm0.03$   & $5.37\pm0.01$ & $-0.48\pm0.03$ &   $-0.42\pm0.02$  \\
 1.6 & $5.35\pm0.02$ &   $5.39\pm0.01$   & $-0.56\pm0.02$   & $-0.61\pm0.01$    	\\
 1.8 & $5.44\pm0.02$   & $5.44\pm0.01$  & $-0.56\pm0.02$   & $-0.61\pm0.01$  		\\
 2.0&  $5.50\pm0.01$ &   $5.48\pm0.01$  &  $-0.71\pm0.02$ &   $-0.82\pm0.01$ 	\\
 2.2 & $5.56\pm0.01$ &   $5.53\pm0.01$  & $-0.91\pm0.01$ &   $-1.03\pm0.01$ \\
2.4 &  $5.61\pm0.01$ &   $5.57\pm0.01$  &  $-1.12\pm0.01$ &   $-1.24\pm0.01$\\
2.6 &  $5.65\pm0.01$ &   $5.61\pm0.01$  &  $-1.31\pm0.01$ &   $-1.41\pm0.01 $\\ \hline	
\end{tabular*}
\end{center}
\caption{Spin-quartet and spin-doublet neutron-proton S-wave
  scattering lengths calculated with ($(a_3^{q})^c$, $(a_3^{d})^c$)
  and without  ($(a_3^{q})^{nc}$, $(a_3^{d})^{nc}$) taking into
  account the topological volume corrections, respectively.}
\label{ScatteringLengthQuartetDoublet}	
\end{table}
The quoted errors reflect the stability of our fits when changing the
number of lattice points and the number or parameters in the fits
(ranging from 3 to 5).

\subsection{Continuum extrapolation of the {\boldmath$nd$} scattering lengths}

\begin{figure}[t]
\centering
\includegraphics[width=1.\textwidth]{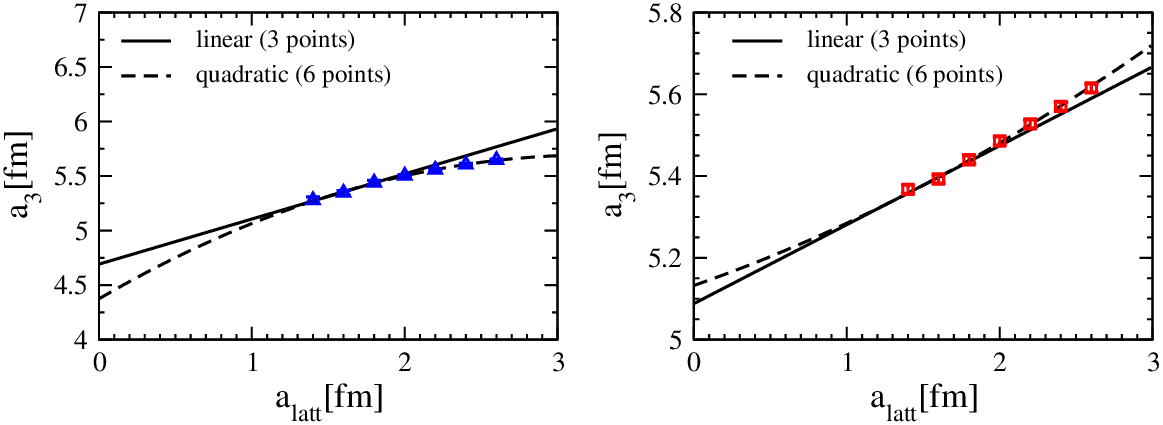}
\caption{Continuum extrapolation for the $nd$ spin-quartet scattering
  length without topological corrections (left panel) and  
with topological corrections being taken into account (right panel).}
\label{qqq}
\end{figure}

\begin{figure}[t]
\centering
\includegraphics[width=1.\textwidth]{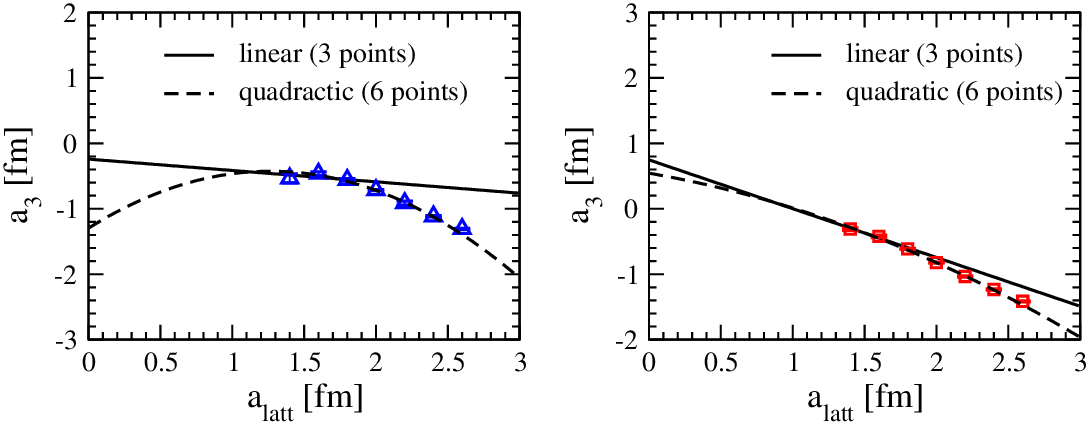}
\caption{Continuum extrapolation for the $nd$ spin-doublet scattering
  length without topological corrections (left panel) and  
with topological corrections being taken into account (right panel).}
\label{ddd1}
\end{figure}

As a  final step in our analysis, we need to perform  a continuum
extrapolation of our results for the scattering lengths obtained at
finite values of the lattice spacing to $a_{\rm latt}=0$. This is achieved
by carrying out a linear and quadratic $\chi^2$ fits to $a_3
(a_{\rm latt})$ at the three
and six smallest values of  $a_{\rm latt}$, respectively. The results of
the fits including the value of $\chi^2$ per degree of freedom are
summarized in Tab.~\ref{quartettable}
\begin{table}[t]
\begin{center}
\begin{tabular}{|c||c|c|c|c|}
\hline
          & Fit       & Points& $\chi^2$ & $a_3$/fm  \\ \hline\hline
without volume    & linear    & $3$     & $0.19$  & $4.69$ \\ \cline{2-5}
 correction & quadratic & $6$     & $0.14$  & $4.37$ \\ \hline\hline
with volume      & linear    & $3$     & $0.55$  & $5.09$ \\ \cline{2-5}
correction   & quadratic & $6$     & $0.22$  &  $5.13$ \\ \hline
\end{tabular}
\hskip 1 true cm
\begin{tabular}{|c||c|c|c|c|}
\hline
          & Fit       & Points& $\chi^2$ & $a_3$/fm  \\ \hline\hline
without volume   & linear    & $3$     & $1.26$  & $-0.24$ \\ \cline{2-5}
correction    & quadratic & $6$     & $1.82$  & $-1.30$ \\ \hline\hline
with volume      & linear    & $3$     & $4.14$  & $0.74$ \\ \cline{2-5}
 correction  & quadratic & $6$     & $3.13$  &  $0.58$ \\ \hline
\end{tabular}
\end{center}
\caption{Continuum extrapolation of the $nd$ scattering length in the
  spin-quartet (left panel) and spin-doublet (right panel) channels.
}
\label{quartettable}	
\end{table}
and visualized in Figs.~\ref{qqq} and  \ref{ddd1}.  
For the spin-quartet channel, we obtain the values 
\beq
( a_3^q)^{nc}=4.53\pm0.16\ \textnormal{fm}, 
\quad \quad  (a_3^q)^c=5.11\pm0.02\ \textnormal{fm}\,,
\eeq
where the quoted error gives the difference between the linear and
quadratic extrapolations. It is, however, likely that the systematic
error associated with the continuum extrapolation is larger than the quoted
uncertainty. Taking into account the topological volume corrections
leads to a clear improvement in the description of $a_3^q$ and brings 
the value extracted on the lattice  in agreement with the one in
Eq.~(\ref{benchq}) obtained in the continuum using the same input.  

In the spin-doublet channel,  after carrying out the continuum
extrapolations the values, we obtain
\beq
( a_3^d)^{nc}=-0.77\pm0.50\ \textnormal{fm}, 
\quad \quad  (a_3^d)^c=0.66\pm0.08\ \textnormal{fm}\,,
\eeq
Here, the effect of the topological volume corrections appears to be
even more pronounced than in the spin-quartet channel. It is
comforting to see that our result for $a_3^d$  after taking into account these
corrections is in a good agreement with the one obtained in 
continuum calculations, Eq.~(\ref{benchd}),  as described in section
\ref{sec3}. This is especially encouraging given the additional
complication in this channel due to the appearance of a near-threshold
pole in the effective range function, which makes the numerical
analysis of the lattice data more challenging.

\section{Summary and conclusions}
\label{sec5}

In this paper we studied the impact of the topological volume corrections
on lattice calculations of low-energy neutron-deuteron scattering. 
More precisely, we used the framework of pionless effective field
theory and restricted ourselves to leading order. In this approach,
the low-energy dynamics of the three-nucleon system is governed by the
Hamiltonian which contains two nucleon-nucleon contact operators 
acting in the $^1S_0$ and $^3S_1$ nucleon-nucleon channels and 
a short-range three-nucleon force which is needed to renormalize
the $nd$ scattering amplitude in the spin-doublet channel. Using a
discretized formulation of the EFT, we determined the values of the
LECs $C_0$ and $C_{I^2}$ accompanying the two-nucleon contact
interactions by the requirement to reproduce the deuteron binding
energy and the $^1S_0$ neutron-proton scattering length. The strength
of the three-nucleon force was tuned to reproduce the triton binding
energy. With the resulting Hamiltonian, we calculated the $nd$
spin-doublet and spin-quartet scattering lengths  $a_3^d$ and
$a_3^q$ with and without taking into account the topological volume
corrections. Throughout our analysis, we employed the values of the lattice spacing parameter in the range 
of $a_{\rm latt} = 1.4 \ldots 2.6$~fm, similar to what is used in
present day chiral EFT nuclear lattice simulations, and
carried out the continuum extrapolation corresponding to the limit of $a_{\rm
  latt} \to 0$. We find that the topological volume corrections play a
substantial role in extracting the $nd$ scattering lengths by changing
the values from $a_3^d=-0.77\pm0.50$~fm ($a_3^q=4.53\pm0.16$~fm)
in the spin-doublet (spin-quartet) channel to $a_3^d=0.66\pm0.08$~fm
($a_3^q=5.11\pm0.02$~fm). Our results agree with the reference
values  $a_3^d \simeq 0.5$~fm and $a_3^q \simeq 5.1$~fm obtained 
in the continuum based on the same input. 
They are also in a satisfactory agreement with the experimental values of 
  $(a_3^d )_{\rm exp}  = 0.65 \pm 0.04$~fm and $(a_3^q )_{\rm
  exp} = 6.35 \pm 0.02$~fm. 
It is particularly encouraging to see that the spin-doublet scattering
length, which is a rather fine-tuned quantity, can be extracted on
the lattice with high accuracy in spite of the additional complication
given by the appearance of a near-threshold pole in the effective
range function. 

The results of our investigation provide an important step towards 
describing nuclear reactions on the lattice, see Ref.~\cite{Rupak:2013aue} for a
recent calculation of the radiative capture process $p(n,
\gamma)d$. \emph{Ab initio} lattice EFT calculations of more complicated
nuclear reactions are expected to become available in the near
future.

\section*{Acknowledgments}

We thank 
Hans Hammer and Gautam Rupak for discussions.
Partial financial support from DFG and NSFC (Sino-German CRC 110), Helmholtz Association 
(contract VH-VI-417), BMBF\ (grant 06BN9006), and U.S. Department of
Energy (DE-FG02-03ER41260) is acknowledged. This work was further supported
by the EU HadronPhysics3 project, and by funds provided by the ERC project 
259218 NUCLEAREFT.


\end{document}